# Digital Twins: Potentials, Ethical Issues, and Limitations

Dirk Helbing[1,2] and Javier Argota Sánchez-Vaquerizo[1]

[1]ETH Zurich, Computational Social Science, Stampfenbachstrasse 48, 8092 Zurich, Switzerland
[2]Complexity Science Hub Vienna, Josefstädter Straße 39, 1080 Vienna, Austria

*After Big Data and Artificial Intelligence (AI), the subject of Digital Twins has emerged as another promising technology, advocated, built, and sold by various IT companies. The approach aims to produce highly realistic models of real systems. In the case of dynamically changing systems, such digital twins would have a life, i.e. they would change their behaviour over time and, in perspective, take decisions like their real counterparts – so the vision. In contrast to animated avatars, however, which only imitate the behaviour of real systems, like deep fakes, digital twins aim to be accurate "digital copies", i.e. "duplicates" of reality, which may interact with reality and with their physical counterparts. This chapter explores, what are possible applications and implications, limitations, and threats.*

## Potentials

Just a few years ago, producing digital twins of dynamical, perhaps even living systems, would have been considered science fiction, and an impossibility from the point of view of science. However, some people believe this situation has recently changed[1] and will do so even more in the future, particularly in view of the incredible amounts of data producible by the Internet of Things (IoT), transmittable by light (LiFi) or other low-latency communication systems, processible by Quantum Computers, and learnable by powerful AI systems. This data collection may be global in scale, but detailed up to the level of individuals and their bodies, using profiling techniques such as those known from social media[2] or even more advanced ones. The upcoming technologies are expanding current personalized services, goods, and devices to the areas of decision-making, behaviour, and health. Due to miniaturization, some components will reach the sub-micrometre scale, as in the case of nanotechnology. Potentially, this enables an Internet of (Bio-)Nano-Things[3] and Internet of Bodies,[4] allowing one to (1) uniquely identify and track everybody via an e-ID, (2) read out data of body functions, and (3) manipulate body functions, supposedly to improve the health of people. It is even conceivable that such powerful digital systems would be used in attempts to maximize "planetary health",[5] considering also the natural resources and ecosystems of planet Earth, while changing human behaviour and civilization as desired by those who control this system.[6]

In the following, we will focus on the limitations and ethical issues of using Digital Twins, while not claiming completeness. Before, however, we want to stress that digital twin technology has certainly uncountable possible applications ranging from production to health, from climate change to sustainability, and from management to politics. It is also often suggested that digital twins will allow one to *predict the future* and *implement optimal control.*[7]

Even though we will question the predictability and controllability of complex dynamical systems, we do not deny the potential benefits of digital twins. They are valuable for their exploratory and prospective power, as they can give an advanced idea of "what-if" scenarios,[8] even when their outcomes are uncertain due to randomness and possible feedback, side, and cascading effects.[9] More importantly, digital twins allow performing otherwise unfeasible experiments. This does not mean, however, that safety precautions are not needed to avoid unethical experiments that would be incompatible with the principles of responsible innovation and engineering. In particular, possible risks for individuals and their environment should be evaluated, and whether they are below an acceptable threshold.

This chapter will discuss digital twins from simple to increasingly complex tasks, highlighting challenges and limitations. We will start with infrastructures and geography, continue with production plants, following up with the environment. Then, we turn to people, phenomena resulting when many people interact, such as traffic flows or stock markets, or cities and societies. Finally, we provide a summary, discussion, and outlook.

## Infrastructures and Geography

Representing *infrastructures* has a long tradition. Starting in previous millennia with sketches and paintings, the age of information technology has brought revolutionary advances, namely in form of helpful software tools for architects, planners, and engineers. Computer Assisted Design (CAD) programs[10] were used widely, and have enabled ever more detailed three-dimensional *visualizations* of planned buildings, allowing for advanced impressions and modifications before they were actually built. These tools enabled the comparison of various building variants, i.e. to assess how a certain building would look like if one did this or that. The concept of *"what-if" scenarios* was widely used with great success. In the meantime, such software tools have been extended into whole Building and City Information Models (BIM and CIM) to *plan* costs, *order* materials just in time, and *manage* the entire construction process, which is becoming more and more complex.[11]

Thus, there is little doubt that digital twins of infrastructures are extremely helpful for planning, construction, and management. A similar thing applies to representations of *geography*. While maps and globes have been used to depict geography for thousands of years, new ubiquitous multi-spectral sensing methods based on *satellite imagery* and pictures taken by unmanned aerial vehicles, often combined with Internet-of-Things and crowd-sourced data,[12] have elevated the field to entirely new levels of detail.[13]

Compared to the systems discussed in the next sections, these applications of digital twins are relatively simple for the following reasons:
- Infrastructures and geographies change very little over time, if at all.
- The underlying structures are material and typically very *well-measurable*.
- More measurements will deliver better data (at least, better averages), if there is not a systematic measurement bias.

Issues arising when other data sets, for example regarding weather, population etc. are overlaid, will be discussed in the following sections.

What is still often missed out in digital twins of infrastructures and geographies are *usage patterns*. However, people frequently occupy space, act, interact, and move around in ways that were not well anticipated when solutions were designed and implemented. This typically makes infrastructures work differently than expected, creating various challenges.[14] We will come back to some of these problems later. Here, we just highlight the ethical problem that infrastructures allow certain people to do certain things, but may exclude others, especially with regard to *ownership, access, and agency*. This implies *competition* and sometimes *conflict, but also inequality* and sometimes *injustice*.[15] Therefore, it is desirable to create *inclusive* spaces, goods, and services (such as public transport, schools, museums, and parks).

From the perspective of architects, engineers, and planners, digital twins may enable true *multi-functional* infrastructures (such as reconfigurable event halls). These allow for *flexible use and adaptation*, which can make usage patterns *less predictable*. While recent technology allows one to measure usage patterns with increasing accuracy, applying "industry 4.0" technologies such as Internet-of-Things-based *sensing* to run "smart buildings", "smart spaces", or "smart cities" implies *privacy* and *security* issues. However, there are further ethical issues. For example, it is conceivable that the *social or medical status* of people would be measured and used to determine access rights to services, care, jobs, or facilities. This could obviously cause new kinds of discrimination.

**Production Plants**

Production plants are particular infrastructures, in which spatio-temporal processes take place, often involving people. Nevertheless, they are typically organized in a *well predictable and controllable* way. *Planning and optimization* are the tools of choice. In fact, this application area was one of the first to use digital twins. Hence, operations research and related disciplines have created a wide range of powerful tools.[16] The *goal (function) underlying the optimization and control of production* is usually clear (classically, it is to maximize profit), and the underlying *utilitarian approach* (which aggregates everything that matters into a one-dimensional index, *score*, or goal function to decide what solution is better or worse) often works reasonably well.

Therefore, digital twins have shown to be very useful for the well-defined tasks of planning, implementation, and management of production.[17] Nevertheless, a detailed representation of *all* relevant processes implies challenges, particularly as production processes may be pretty sophisticated and complicated. This includes the following:
- Some optimization and control problems are *NP-hard* (i.e. computationally complex, requiring a lot of computational resources).[18] Therefore, the optimal solution cannot be found and implemented in real-time. It must be determined offline, or *simplifications* must be made, which might reduce the system performance (see also the section on traffic control).
- *Delays* between requiring to make adjustments (e.g. of the quantity produced to meet market demands), taking the related decisions, implementing them, and becoming effective, may cause *instabilities*. This may imply *limited predictability and control*.[19]
- Elements of a production plant may fail, machines may break down, or supplies may fall short (see, for example, the recent chip crisis). There may

- also be blackouts of electricity or communication or cyberattacks. All of this further reduces predictability and control.
- It may be difficult to break down costs and estimate or measure *side effects (externalities)*, which may imply unexpected costs (such as bankruptcies of suppliers, economic crises, lockdowns, or legal cases).

The last point raises particular ethical concerns, as the neglect of *externalities* in the past has caused relentless resource exploitation and serious environmental damage, climate emergency as well as social and health issues, to mention just a few problems. Therefore, carbon taxes shall now be introduced in order to "price in" the damages caused by $CO_2$ and to incentivize carbon reduction. Overall, however, it has been concluded that measuring economic success by one index such as profit or gross domestic product per capita, is insufficient, such that multiple indices are needed to measure success.[20] This, however, questions the suitability of the utilitarian approach and related optimization methods based on it, particularly when a societal perspective is taken, or nature is considered.[21]

## Environment (Climate, Weather, Ecosystems, Nature)

Recently, it has been proposed to build digital twins also of environmental and climate systems.[22] As the climate depends on consumption and emission patterns, such a digital twin would probably consider – at least to some extent – models of the world economy and societies on our planet. Some experts, therefore, demand that such digital twins should consider the effects of production and human activities on nature, and the effects of the environment on humans, society, and production.[23] This implies further challenges (we will first focus on the environment, here, while the challenges of capturing and simulating socio-economic dynamics will be addressed later):

- The accurate representation of the environment requires a massive amount of data. *Remote sensing* (e.g. by means of satellites) may not be enough to get the best possible representation. Therefore, *ubiquitous sensing*,[24] using the Internet of Things, has been proposed, which would involve millions or billions of sensors, or even a lot more, if nanotechnology or "smart dust"[25] would be used as well.
- Measurements are never exact. There is always a finite *confidence interval. Context* and *interpretation* may matter.
- There are many *issues with Big Data analytics*. These include sensitivity, false positives/negatives, biases and discrimination effects, overfitting, spurious correlations, difficulties to distinguish causation from correlations, etc.[26]
- Random, *probabilistic effects* (sometimes called "noise") may also play an important role.
- There are *fundamental limits* to predictability due to "chaos" and "turbulence" (see the example of weather forecasts[27]), undecidability (see the theorem of Kurt Gödel[28]), and computability (NP-hard problems), plus fundamental physical limits (e.g. due to the uncertainty relation).
- The *convergence* of "black box" learning algorithms (for example, neural networks, or deep learning) implies additional issues:[29] machine learning is an iterative process, which takes time and assumes convergence to the truth. However, reality might change more quickly than it takes to learn. This problem is amplified by the following points.

- In contrast to simple, linearly interacting systems, environmental systems often display *complex dynamics*.[30] This implies, for example, emergent properties and phenomena (and often counter-intuitive, surprising behaviours),[31] unexpected sudden phase transitions at "tipping points"[32] (often called "catastrophes"[33]), power laws implying extremely fluctuating statistics,[34] and overall, limited predictability[35] and controllability,[36] even if a lot of accurate measurement data are available.
- *Network interactions* may further increase the previous issues and add undesired *side, feedback, and/or cascading effects*. Combined with probabilistic effects, this may cause *uncertainty*, vulnerabilities, and systemic risks.[37]
- Many *wicked problems*, where measures taken to solve a problem produce even bigger problems, are a result of complex dynamics in networked systems.[38] Nevertheless, complex dynamical systems should not be considered a bad thing. They are not. However, a successful understanding and management of complex dynamical systems require a paradigm change, namely a shift of the focus from the (material) system components to their (typically immaterial, invisible, and hardly measurable) interactions.

Overall, when it comes to complex dynamical environmental phenomena such as weather, climate, or ecosystems, the representation accuracy of digital twins is limited, not just by the amount of data one can measure, process, and transmit. On top of this problem comes the fact that the *data volume* grows faster than the *processing power*. Moreover, a *data-driven approach* based on Big Data analytics often does not fulfil the requirements of solid *data science*, which typically requires domain knowledge and experimental tests using proper statistical methods. That is, "data-based" must be distinguished from "fact-based" (as the latter requires the use of *verification and falsification* measures).

All of this can lead to errors and misrepresentations as well as wrong courses of action.[39] As a consequence, a new approach called "global systems science" has been proposed.[40] While the term "evidence-based" has mostly been used for "fact-based" in the past, it is now increasingly being used for "data-based", which is confusing and problematic, exactly because there are fundamental limits to the accuracy of digital twins in environmental complex systems (and others discussed in the following). The problem is further amplified if policies are taken in response to extrapolated or inferred data, or in response to *predicted* futures, which may never materialize. In addition, there is often a gap between scientifically suggested responses and policy responses, and a gap between intended and actual solutions.

Of course, from an ethical point of view, there may be further ethical issues. For example, the *surveillance* of consumption or other behaviours implies *privacy* issues. A *scoring* of people may imply discrimination, and *targeting* them may undermine their freedom rights, particularly when behavioural manipulation is involved.[41] All of this can be in conflict with *human dignity*.

**Traffic Flows and Markets**

Traffic flows and financial markets constitute two other kinds of complex dynamical systems, this time involving large numbers of people. Let us start with traffic flows.[42] Here, one can find a range of diverse congestion patterns, which depend on the traffic volume and inhomogeneities along the road (such as on- and off-ramps, gradients, or changes in the number of lanes).[43] Interestingly, due to the interactions of many vehicles, traffic flows are multi-stable, i.e. even for identical traffic conditions, there can be different outcomes, depending on local fluctuations in the density. In other words, traffic flows display the feature of history dependence (*"hysteresis"*), and there are *counter-intuitive* phenomena such as "faster-is-slower effects" (i.e. trying to go faster may trigger a breakdown of the traffic flow, which slows down traffic).[44]

These congestion patterns are a consequence of *systemic instability*:[45] above a certain critical density, random variations in the traffic flow are being amplified, causing a *domino effect*. It is striking that particular kinds of adaptive cruise control systems can dissolve undesirable collective effects such as traffic jams without the need for centralized control, just based on local self-organization.[46] Here, the real-time measurement of the distance and relative speed of the vehicle ahead is used by a particular adaptive cruise control system for real-time feedback to slightly change the acceleration or deceleration of a vehicle. Based on such "*distributed control*",[47] traffic jams can be dissolved by changing vehicle interactions in a minimally invasive way as compared to what drivers would do. The digitally assisted, local self-organization leads to another, desirable, collective outcome: free traffic. So, real-time feedback and coordination approaches do not only work in complex systems with many dynamically changing variables – they may even be more successful than classical control approaches. A similar finding has been made for traffic light control.[48]

Financial markets are another interesting example to discuss. Here, one important finding is that *markets would not work, if everyone decided and behaved according to the same deterministic rules*. Surprisingly, markets get more efficient, if some degree of *heterogeneity* or "*noise*" is involved, i.e. if the actors decide in diverse ways. A model to illustrate some underlying principles is the so-called "minority game".[49] In this game, majority decisions tend to be less profitable than minority decisions. That is, better results can be achieved by betting against the majority.

Both examples, traffic flows and markets, illustrate that self-organization in complex systems should not be considered a problem, but can be a good *solution*. For this, having the right kinds of interactions among the system components (here, drivers or traders) is more important than the individual properties of the system components.

Finding suitable interactions, which will automatically and efficiently produce desirable outcomes based on self-organization, is a matter of *"mechanism design"*[50] and *"complexity science"*.[51] A favourable side effect of such self-organization is that the resulting collective behaviour of the system is typically robust to reasonably small perturbations ("disruptions"), which implies systemic *resilience*.[52] However, lack of transparency and required real-time data, wrong feedback and incentives, or lack of accountability and coordination might lead to *systemic failures*,[53] as urban gridlock or financial crashes illustrate. Therefore, self-organization (sometimes also called the "invisible hand"[54]) does not *always* deliver desirable outcomes, if everyone just does whatever they like.[55] As pointed out before, one needs suitable interactions.

Importantly, however, self-organization allows for some degree of *freedom* (usually without much loss of functionality or efficiency)*,* which does not usually apply to centralized control.[56]

Faced with undesirable *"misbehaviour"*[57] of complex systems, one often calls for more control of their system components. Accordingly, everybody should show a certain, desired behaviour. However, above we have learned that this can be counterproductive. *Diversity* is frequently needed for techno-socio-economic systems to work well[58] and, accordingly, digital twins need to be diverse (not only in terms of individual parameter specifications but also with regard to their inner working, i.e. their underlying algorithm).

Moreover, centralized control attempts tend to perform poorly or fail in systems, which are characterized by randomness, variability, heterogeneity, network effects, and complex dynamics, where internal interactions typically prevail and predictability is often limited. In these systems, a flexible and prompt adaptation to the respective local conditions and needs is a promising approach to promote coordination and favourable self-organization in the system, while the possibility of optimal control is often an illusion.

Despite the specificity of traffic flows and markets, similar complexity-related challenges are abundant in many cities around the world.[59] To accomplish fully functional, real-time, and bi-directional physical-virtual frameworks able to manage mobility and other complexity challenges effectively,[60] further research is required.[61] Recent research focuses, for example, on connecting mobility with heterogeneous socio-economic interactions and urban logistics.[62] Reflecting the diversity of actions and preferences in reality precisely is one of the goals of digital twins that increasingly aim to mirror entire economic systems.[63] Starting from operations for business intelligence,[64] they ultimately strive to optimize markets and financial ecosystems for more sustainable development[65] or other goals.

**Precision Health**

Digital twins have been proposed also for use in medicine and health care.[66] For example, they have been applied to prepare for difficult surgeries.[67] Eventually, digital twins are expected to capture body structures, functions, and processes not only on a macro-scale, i.e. for organs[68] or entire bodies,[69] but also on a micro-scale, i.e. on a cellular level or below.[70] This requires entirely new measurement methods. Nanotechnology has been proposed to offer novel solutions: exposing the body to nanoparticles or nanobots may allow one to read out activity patterns and realize new approaches to health care.[71] Relevant technologies have become known under names such as "Internet of (Bio-)Nano-Things",[72] "Internet of Bodies",[73] or "Internet of Humans".[74] The harvested data may be used for "personalized medicine" or "precision health".[75] Proposed applications might range from fighting cancer to brain activity mapping.[76] Ray Kurzweil and others even suggest we might be able to overcome death and, thereby, live forever.[77]

However, new challenges arise, such as the following: 1. Is life extension in an "over-populated world with limited resources" ethical, or would it amplify inequality and shorten the lives of some people for the benefit of others? 2. How to avoid technological, social, behavioural, or eugenic selection, [78] which could mean bias, discrimination, and population control violating the right to life? 3. What about the health threats resulting from the limited accuracy of "read" and "write" operations in the sub-millimetre range? 4. How to avoid dual-use, when it becomes possible to read minds remotely and to engineer thoughts, emotions, values, decisions, and behaviours?[79] 5. How to protect ourselves from theft or manipulation of our highly sensitive health data in view of cybersecurity issues and hacking threats?[80] 6. How dangerous for our well-being would it be, if two operating systems interfered with each other: the natural operating system of our body and a digital, data-driven, AI-controlled one? A discussion of such and further issues of a Planetary Health agenda is available elsewhere.[81]

## Humans (vs. Robots)

Traffic and stock markets are not where complexity ends. People are (to a large extent) complex self-organizing systems themselves: many body functions, including brain functions, are complex and adaptive to a changing environment.[82] The fact that people have a brain implies a number of additional challenges that make it even more difficult to produce highly accurate digital twins:

- *Information processing* comes into play and has a dominating role in the resulting individual behaviour.
- *Decision-making* becomes important. While companies may strive to choose the best performing solution, which often narrows down the number of relevant options to one, people care about having *alternatives and freedom of choice*. Typically, human decisions are not based on strict optimization, but on heuristics,[83] e.g. simple decision rules such as the principle of "satisficing"[84] (i.e. choosing an option that is considered to be "good enough").
- People *learn*, such that their way of decision-making changes over time.
- People exchange information, for example, by using *languages*. In this connection, it is important to know that languages are not fixed and given for all times, but complex, adaptive, and evolving systems themselves.[85]
- People can give things a *meaning*. In fact, all words become meanings by shared usage patterns, called "conventions".[86]
- Language is often *ambiguous* and only fully understandable in context.
- People have *consciousness*, which until today has been conceptualized and understood only to a limited extent.
- They often act according to individual *intentions*.
- Their *goals change* over time (e.g. if they are hungry, they will look for food, but after eating, they will turn their *attention* to something else). Accordingly, their behaviour cannot be well understood by means of one utility function, particularly one that does not change.
- People act in many different contexts that are characterized by different *norms* (i.e. commonly expected behaviours). Accordingly, they play many different *roles*, which are to be modelled by different *personas*.
- They have a *self-image*, which guides their own behaviour, and *images* of others, which determine their expectations and actions.
- Most people have *empathy*, which may trigger other-regarding behaviour.
- People have *emotions*, which change their consideration and valuation of options, among which they may decide.

- People may *feel pain*.
- People are *playful*, i.e. they do things that do not have an immediate purpose apart from having fun.

The above findings pose particular challenges for creating digital twins. The discipline of *cognitive computing* is trying to account for (some of) these features, but requires a massive amount of sensitive personal data to create increasingly detailed digital twins,[87] which obviously raises *privacy* issues. However, no matter how much data is available, it is highly doubtful that one could ever create an identical digital twin. Many of the relevant variables are not directly measurable or measured – and must, therefore, be *inferred* from other observations.[88] For example, diseases are often inferred from symptoms, intentions from actions, and meanings from context. This implies the risk of *misinterpretations*, as it is also known from AI-based emotion classifications using facial expressions.[89] The possibility of ambiguity and mimicry or deception must be considered as well.

It must further be stressed that modelling humans, their thinking, feeling, decisions, and actions do not only concern the area of brain science, but also the social sciences and humanities, which tend to doubt that all of the above concepts can be quantified and measured well, or operationalized in a way that could be translated into algorithms. For example, people particularly care about non-material qualities such as *consciousness, dignity, creativity, friendship, trust, and love*, which are not well measurable (or even hard to define), and will perhaps never be quantifiable in an adequate way. There is certainly a danger to ignore human features that are not represented by numbers, or perhaps not even representable by numbers at all. Accordingly, a data-driven society may easily neglect important qualities, which would possibly produce an *inhumane society*.

From the point of view of social science, humanities, and law, treating people just like things or robots (that may be reset, replaced, used or changed in arbitrary ways, or thrown away) would be considered highly inappropriate and *unethical*. It would violate their *human dignity*. It would also undermine their ability to self-organize. It would further affect their *autonomy* and *freedom*. We would like to underline, here, that freedom is not primarily a matter of allowing for selfish behaviour. It is a precondition for experimentation, learning, and innovation. It is also important to enable the variability needed to make the many different roles compatible, which people are trying to fulfil every day.

## (Smart) Cities and Societies

A further level of complexity (implying additional challenges to produce accurate digital twins) is expected for cities and societies, particularly "smart cities" and "smart nations".[90]

- While smart cities and smart nations are delivering a lot more data than was ever imagined in the past, with the increasing amount of networking and interconnectivity, *the level of complexity is growing even faster than the volume of data* (namely, in a combinatorial, i.e. factorial rather than exponential way).[91] Paradoxically, this can cause an increasing loss of (centralized) control, even though we have more data and better technology than ever (which calls for distributed control approaches that can support a favourable self-organization of the system).

- Social systems are *multi-level complex systems*. People, for example, who are complex systems themselves, create complex social systems based on self-organization and emergence. Group formation, for instance, will come with an emergent *group identity*, which will change the behaviour of the group members.[92] Note that this is very different from atoms that form molecules (which will usually not change the properties of the atoms in a significant way).
- Given the many different social contexts people are part of – and the associated different roles expected from them – people have multiple and changing goals. Therefore, cities and societies are faced with a *plurality of goals* that need to be met. Accordingly, they should not be run like production plants or companies, which traditionally serve one goal, such as profit maximization. Politics is needed to bring the many different goals into a suitable balance, which may, of course, change over time. Accordingly, the diverse goals of a city are typically not well represented by a utility function.
- The network interactions of people create *social capital* such as trust, reputation, or solidarity.[93] While this is of great value for the economy and society, it has been difficult so far to operationalize and measure social capital.
- To a much greater extent, the same quantification problem exists for *culture*. Defining culture as a collection of social norms and routines or success principles seems to capture only some of its essence. For example, culture also has to do with the playful nature of humans, with the exploration of uncharted waters, with learning and innovation.
- Societal decisions are not just about optimization and control. They trigger the exploration of new solutions and result from interactions in many parts of the system. Therefore, *co-evolution* is a much more adequate description of what characterizes cities and societies.[94]

It is questionable whether all of the above is already being considered, or even *can* be considered, by digital twins. Attempts to build digital twins of entire societies exist at least since the Sentient World Project.[95] Promoted by the US Department of Defense and various Fortune 500 companies,[96] this simulation platform has also been used for planning wars and population-scale psychological operations (PsyOps).[97] It is not far-fetched to assume that this or a similar platform has been developed further in order to create a digital twin of the world, based on mass surveillance data.[98]

Recently, it has even been proposed that considering detailed knowledge about everyone's opinions and preferences would allow one to create a democratic post-choice, post-voting society.[99] Mass surveillance could create entirely new opportunities, here. The underlying *technocracy* would automatically aggregate opinions, while politicians would no anymore be needed to figure out and implement what people want and need.[100] Rather than an upgraded democracy, however, such a society could become a novel kind of *digital populism*, in which the will of majorities might be relentlessly imposed on everyone, thereby undermining the protection of minorities and diversity.

As it is more efficient to deal with data rather than people, there is also the danger that our digital twins would be given *greater authority* than us, even though the representation of our will and us by our digital twin could be *inaccurate, biased, manipulated, or hacked*.[101] That is, if there is a disagreement between a person and the corresponding digital twin, the system would assume the data of the digital twin and ignore the opinion of the human it should represent.

Therefore, rather than replacing individual choices with automated machine decisions, we recommend going for digital assistance of decision-making, offering people individually and systemically good opportunities. That is, instead of identifying, taking, and implementing *one* optimal solution to a given goal (e.g. taking the shortest route), the decision-support system should offer a number of high-performance solutions, and for each of them, it should indicate a diverse set of qualities relating to various relevant goals. For example, if I need to go from A to B, and one route takes 102 minutes, while the other takes 103 minutes, there is no reason to force me to take the faster route. If the other route is more scenic and I am not in a rush, it would make sense to take the slightly longer route. This would probably have positive side effects that are well-known, but hardly quantifiable, such as having a better mood when I arrive, which will lead to more agreeable decisions, a better team spirit at work, and higher creativity of everyone. The longer route might also come with less $CO_2$ emissions (if used at a lower speed), or it may allow me to do some shopping along the way, which will save time and emissions later during the week. Therefore, "scenic", "shopping", and "fewer emissions" would be some of the relevant qualities of high-performance solutions, when it comes to choosing optimal routes. While offering such choices would make the system perhaps less predictable in detail, it is likely that it will improve the state of the world in many hardly quantifiable aspects, which will cause further combinatorial benefits to society.

In contrast, running a society based on a digital twin can have undesired side effects, if not done wisely. It might cause *lock-in effects* in potentially outdated operation principles of the past. It might also promote a society, which is too much oriented toward control rather than creating opportunities. Moreover, a digital twin of society, which includes detailed digital twins of its people, could be easily abused. For example, by knowing the strengths and weaknesses of everyone, one could trick or manipulate everybody more effectively,[102] or mob them with hate speech on social media. Furthermore, a digital twin of society would also make it possible to determine how much one can pressure people without triggering a revolution, or figure out how to overcome majorities, how to break the will of people, and how to impose policies on them, which do not represent their will. Such applications of mass surveillance might be considered to be highly *parasitic* and *undermine human rights*.

Even if it would not come this bad, with the growing power of information technology, there is certainly a risk that methods originally developed for the management of supply chains, companies, or theme parks would be transferred to cities and societies, without realizing the *categorical mistakes* made by this. Moreover, such an approach may be even ignorant about the many (over-)simplifications made, neglecting details and hardly measurable aspects (such as human dignity), which would be treated like "noise". Such an approach could destroy the main strengths of social systems: their ability to innovate and adapt, to self-organize, and (co-)evolve. In fact, it might even *destroy societies* as we know them, just for the sake of more control.

It is to be expected that, for many people, this would not end well. There is definitely a danger that running societies in a data-driven and AI-controlled way could lead to an inhumane organization of people, which some people would characterize as *"technological totalitarianism"*. But what should we run society for, if not for the people?

## Summary

We would like to summarize this chapter with 12 statements on digital twins:[103]

1. **On Data:** It has become an attractive idea to create digital twins of everything, including the Earth, its climate, human bodies, and their health. While this approach may have many benefits, there are limits. All in all, one must realize that a data science rather than a merely data-driven approach is needed, which requires sharing a lot more data with a lot more people.

2. **On Complexity:** Creating an accurate digital twin for infrastructures, which change little over time, is easy. However, it will probably never be possible to produce an *exact* digital twin of life on Earth, even if nanotechnology is being used for ubiquitous sensor measurements. One is faced with fundamental challenges and measurement limits when models of complex dynamical systems are being built, for example, of weather, climate, or life, of brains, behaviours, or health. Thus, one needs to be prepared for uncertainty.

3. **On Machine Learning:** The biggest publicly known modern machine learning models try to learn a trillion parameters or so. Unpublished corporate, governmental, or military models may contain even more parameters. While this is impressive, more predictive power is often achieved by simpler models (think of "over-fitting"). Surprisingly, noisy or little data can sometimes generate better models.[104] But no matter how many variables are being considered, there are many orders of magnitudes of interaction effects that are not captured, hence neglected. This can produce a wrong picture and bad forecasts, which can be dangerous.

4. **On Artificial Intelligence:** So far, Big Data has not been able to replace science – in contrast to what Chris Anderson had envisioned,[105] nor do we have a universal AI/Artificial General Intelligence (AGI), according to what is publicly known. Even if we had one, this could still be dangerous, particularly if not retaining meaningful human control.[106] Suppose, for example, one would task an intelligent system to solve the sustainability problems of the Earth or to maximize planetary health. This might result in depopulation and trigger an 'apocalyptic' scenario, even though a better future for everyone might exist. Moreover, as many of today's AI systems operate like "black boxes", one might not even realize some of the harmful effects AI systems are causing.

5. **On Optimization:** The concept of "optimizing the world" is highly problematic because there is no science that could tell us what is the right goal function to choose: should it be GDP per capita or sustainability, life expectancy, health, or quality of life? The problem is that optimization tries to map the complexity of the world to a one-dimensional function. This leads to gross oversimplifications and to the neglect of secondary goals, which is likely to cause other problems in the future. Using (co-)evolutionary approaches would probably be better than optimizing for one goal function. Coordination approaches may be more successful than control approaches.

6. **On Qualities:** A largely data-driven society is expected to perform poorly with regard to many hardly measurable qualities that humans care about. This includes freedom, love, and creativity, friendship, meaning, dignity, and culture, in short: quality of life...

7. **On Innovation:** Something like a "digital crystal ball" is unlikely to see disruptive innovations, which are not included in the data of the past. Hence, predictions could be too pessimistic and misleading. For example, consider the forecast of the world population. According to some future projections, about one-third of the world's population is sometimes claimed to be "overpopulation". Consequently, these people may get in trouble when managing the world via a digital twin, as its projections do not consider better ways of running our economy, which may be invented. Probably, "overpopulation" is not the main problem, but lack of economic (re-)organization.

8. **Humans vs. Things:** In a highly networked, complex world, where almost everything has feedback, side, or cascading effects, ethical challenges abound. For example, people should *not* be managed like things. In times when many argue about "trolley problems" and "lesser evils", if there is just a big enough disaster, problem, or threat, any ethical principle or law may be overruled, including human rights and even the right to life. Such developments can end with crimes against humanity.

9. **On Dual Use:** A powerful tool, particularly when applied on a global scale, may cause serious, large-scale damage. It is, therefore, necessary to map out undesired side effects of technologies and their use. Effective measures must be taken to prevent large-scale accidents and dual-use. Among others, this calls for decentralized data storage, distributed control, and quality standards. Moreover, transparency of and accountability for the use of data and algorithms must be dramatically improved, and participatory governance should be enabled.

10. **On Alternatives:** We should carefully consider alternative uses of technology. Here, we would just like to mention the idea of creating a *socio-ecological finance system*: a finance system, which would use the Internet of Things to measure externalities that decisions of people and companies cause.[107] The measurement of externalities would define multiple new currencies, which could locally incentivize positive behavioural change. This novel real-time feedback and coordination system is inspired by nature. Nature has already managed to develop a circular economy based on self-organization and distributed control. Hence, introducing real-time feedback into our socio-economic system could create forces promoting a sustainable re-organization. A *sustainable circular and sharing economy* would result through a co-evolutionary process. If designed in a value-sensitive way, it could be a system consistent with freedom, privacy, and self-determination, with creativity and innovation, with human rights and democracy. This would probably be the best path to sustainability currently known.

11. **On Governance:** As people are increasingly an integral part of socio-technical systems, a technology-driven approach and technological innovation are not enough. We first and foremost need *social innovation* to unlock the benefits of the digital age for everyone. A platform supporting true informational self-determination is urgently needed. Moreover, the classical *war room* approach needs to be replaced by a *peace room* approach, which requires, among others, an interdisciplinary, ethical, multi-perspective approach, in other words, a new multi-stakeholder approach to achieve better insights and participatory resilience.[108]

12. **In Conclusion:** Societies are not machines, and an optimization approach is too narrow to manage them.[109] It is, therefore, important to recognize that complexity is an opportunity for new kinds of solutions, not "the enemy". Planning should be increasingly replaced by flexible adaptation, optimization by co-evolution, and control by coordination. Obviously, all of this can be supported by *digital assistance*, if used wisely and well.[110]

## Outlook: From the Metaverse to "The Matrix"?

Many readers may have noticed the re-birth and extension of the cyberpunks' virtual world, advocating a "Second Life", in the recently rebranded "metaverse". The massive investments imply bigger plans. The metaverse is not just a "parallel world". It likes to offer more than opportunities to create accurate mirrors of real, physical systems, using privacy-invasive techniques such as tracking, profiling, or scanning humans (including, for example, hand gestures, iris movements, and dilation)[111], or even applying nanotechnology. Going a step further, the metaverse even wants to enable real-time feedback between the digital and physical realms.[112]

What we are seeing now is just the beginning. As long as real bidirectional interactions are lacking, it could be argued that digital twins are not yet fully functional. To accomplish the full vision of digital twins, they must be able to actuate and modify the physical environment they are mirroring, but that could sometimes go wrong.

In any case, progress is quick. Imagine, for example, that it will become cheap and common to use augmented reality glasses, or to get a more realistic embodiment experience, or to project digital twins into our physical environment (by means of hologram technology) or our minds (by means of neurotechnology – or even to materialize our avatars in form of robots (if not androids). This would make digital twins – and the people using them – a lot more powerful. For example, people could interact with others remotely, without requiring transportation. This could reduce the severe environmental impact of transportation, enable access to dangerous environments, and expand the capacity to interact with people without spatial limitations, allowing them to work, study and play together without the constraints of our physical world.

Yet, new ethical questions would arise. For instance, one might try to commodify digital twins and assets in metaverse-like virtual environments,[113] enabled by pervasive monitoring of individual behaviour at a level of detail that is inconceivable in the physical world, but completely feasible in immersive worlds.[114] Also, new forms of identity theft, abuse, and deception would be possible. For example, how to reach a fair society, when rich people can "multiply" themselves, being represented in parallel by multiple avatars or robots? Furthermore, how to make sure that the limited resources of planet Earth will not be wasted on unnecessary technology, which tempts people to live in an illusionary virtual world, escaping the real problems of the physical world,[115] rather than using them in favour of humans and nature? Last but not least, how to prevent we will end up in "The Matrix" – a world where people would be bounded by digital technologies? And how to prevent people will be entirely replaced by digital twins in an extreme form of transhumanism?[116] The list of potential issues could certainly be expanded...

**Conclusions**

Digital twins of all sorts of systems are now becoming very popular. While many successful applications are already known in architecture, geography, manufacturing, and logistics, applications to traffic and smart cities are on the way. There are also attempts to build digital twins of human bodies and health, of people and society. However, they are faced with measurement and big data challenges, extreme levels of complexity, and ethical issues. Applications that interfere with individual thoughts, decisions, behaviours, and bodies are particularly problematic. Therefore, it is concerning that many strategy papers do not stress human rights and human dignity, or do ignore them altogether.

It is conceivable, for example, that digital twins would be built, using a nanoparticle-based measurement of body functions ("in-body surveillance").[117] While this may create opportunities for entirely new health services, it may also enable body hacking and the stealing of highly sensitive personal health data. Hence, besides many new opportunities, new kinds of cybercrime may arise, including forms of identity theft that are probably hard to uncover and defend against. This development is dangerous. It could potentially lead to a "militarization" of everything from food, air and medicine to bodies and minds.[118] Therefore, apart from technological innovations, social innovations are required to manage them. To unlock the full potential of new technologies, novel, participatory governance frameworks are required, which give particular weight to a scientific approach and to the individuals affected by any (potential) measures taken, considering relevant alternatives. In order to make sure applications will be beneficial and benevolent, a particular focus on the most vulnerable is recommended. Most likely, we will need a *new social contract*. This would have to include the possibility of self-managing sensitive personal data. It is still to be seen whether Web 3.0 will deliver an appropriate and well-working solution for this.

If these problems are not solved soon, however, politics and citizens may lose their birthright to co-create their future. Currently, we seem to lack sufficient protection mechanisms to prevent that, for example, pervasive mass surveillance, repressive opinion control, a restrictive cashless society, and/or neurocapitalism[119] could be put in place at some point in time, even without full and well-informed consent of people.[120]

It further appears that we do not have enough protection to prevent the use of sensitive personal data (particularly behavioural and health data) against us. This is a serious threat that calls for novel, participatory governance approaches,[121] which can minimize the misuse of powerful digital technologies while maximizing benefits for everyone.[122]

There are better ways to use technologies for a more sustainable and healthier world, in harmony with nature,[123] than creating something like a technocracy, run from a war room that lacks sufficient consideration of privacy, ethics, transparency, participation, democracy, and human rights.[124] Planetary health and human well-being cannot be reduced to a problem of optimization, supply chain control, or resource management. Doing so would dramatically oversimplify the world and fall short in view of complex systems, their various interaction effects, and emergent properties.[125] It will also miss out on the opportunities that complexity offers, such as self-organization, emergence, and co-evolution.

Rather than aiming for perfect digital twins, a predictable future, and total control, one should use computer simulation technology to create better opportunities for everybody, e.g. for digital assistance of creativity, innovation, self-organization, coordination, cooperation, and co-evolution. The metaverse, for instance, if used well, could provide a helpful experimental playground that would allow one to try out alternative organizations of cities and societies. In this connection, participatory formats inspired by the classical *Agora* are particularly appealing.[126] They are also promising for developing participatory approaches to achieve higher resilience.[127]

Last but not least, digital technology allows us now to develop entirely new solutions that are based on a distributed, flexible adaptation to local needs, on digitally assisted self-organization, and co-evolution. This may come with less predictability and control, but is expected to improve the sustainability and carrying capacity of our world, while promoting quality of life, prosperity, and peace.[128]

## Acknowledgements

JASV is grateful for partial financial support from the Future Cities Lab Global at the Singapore-ETH Centre, which was established collaboratively between ETH Zurich and the National Research Foundation Singapore.

DH acknowledges the excellent research opportunities within the project "CoCi: Co-Evolving City Life", which receives funding from the European Research Council (ERC) under the European Union's Horizon 2020 research and innovation programme under grant agreement No. 833168.